\documentclass{article}

\usepackage{arxiv}

\usepackage[utf8]{inputenc} 
\usepackage[T1]{fontenc}    
\usepackage{hyperref}       
\usepackage{url}            
\usepackage{booktabs}       
\usepackage{amsfonts}       
\usepackage{amsmath}
\usepackage{amssymb}
\usepackage{nicefrac}       
\usepackage{microtype}      
\usepackage{graphicx}
\usepackage{xcolor}
\usepackage{subcaption}
\usepackage[numbers]{natbib}

\graphicspath{ {./Figs/} }

\title{A Learning-Based Method of Automatic Music Transcription}

\author{
  Phu-Anh-Tai Do\thanks{Equal contribution.} \\
  Department of Information Technology \\
  FPT University \\
  Ho Chi Minh City, Vietnam \\
  \texttt{} \\
  \And
  Huu-Hoang Nguyen\footnotemark[1] \\
  Department of Information Technology \\
  FPT University \\
  Ho Chi Minh City, Vietnam \\
  \texttt{} \\
  \And
  Duc-Hoang Nguyen\footnotemark[1] \\
  Department of Information Technology \\
  FPT University \\
  Ho Chi Minh City, Vietnam \\
  \texttt{} \\
  \And
  Thanh-Hai Le\thanks{Corresponding author / Project Supervisor.} \\
  Faculty of Information Technology \\
  Ho Chi Minh City Open University \\
  Ho Chi Minh City, Vietnam \\
  \texttt{} \\
}

\begin{document}
\maketitle

\begin{abstract}
Automatic Music Transcription (AMT) seeks to convert acoustic music recordings into symbolic musical notation, a process that remains remarkably difficult for polyphonic piano due to concurrent polyphony, harmonic partial interference, expressive timing, and severe pitch activation sparsity. In this paper, we formulate polyphonic piano transcription as a frame-level multi-label classification problem across the standard 88 piano pitches (MIDI notes 21 to 108). Raw audio waveforms are processed using the Constant-Q Transform (CQT) to construct a geometrically spaced log-frequency time-frequency representation that aligns with the chromatic musical scale across seven octaves (252 frequency bins per frame). We systematically investigate recurrent sequence models, contrasting unidirectional Long Short-Term Memory (LSTM) with Bidirectional LSTM (BiLSTM) backbones, and propose two architectural adaptations tailored to recurrent temporal representations: a modified additive attention module and a modified multihead attention mechanism equipped with hyperbolic tangent activation and head-wise summation. Furthermore, to mitigate the severe class imbalance between inactive and active notes (approximately 41:1), we evaluate Weighted Binary Cross-Entropy (WBCE) alongside standard Binary Cross-Entropy (BCE). Rigorous empirical evaluations on the MAPS and MAESTRO datasets demonstrate steady, compounding gains from bidirectionality and attention mechanisms. Our proposed Multihead Attention--BiLSTM architecture attains an F1-score of 88.20\%, an accuracy of 78.90\%, a precision of 0.89, and a recall of 0.86, substantially outperforming the baseline unidirectional LSTM (F1-score of 71.20\%) while maintaining a compact parameter budget of 4.43M. Qualitative piano-roll visualizations confirm that our model generates faithful transcriptions even in complex harmonic passages.
\end{abstract}


\section{Introduction}
\label{sec:intro}

The transcription of acoustic music into written symbolic notation represents a remarkable feat of human intelligence, uniting psychoacoustic perception (auditory scene analysis), cognition (auditory object identification), formal knowledge representation (musical syntax and harmony), and probabilistic inference (evaluating competing musical hypotheses) \citep{benetos2018automatic}. Automatic Music Transcription (AMT) aims to automate this complex translation process by designing computational algorithms capable of converting acoustic signals into symbolic representations such as MIDI or digital sheet scores. AMT constitutes a cornerstone of Music Information Retrieval (MIR), offering foundational primitives for downstream applications including musicological analysis, audio search, automated accompaniment, cover-song identification, and computational music education \citep{klapuri2006, benetos2018automatic}.

Despite decades of active research, polyphonic AMT---and polyphonic piano transcription in particular---remains an inherently ill-posed and challenging problem. This difficulty arises from four primary acoustic and structural phenomena:
\begin{enumerate}
    \item \textbf{Polyphony and Overlapping Mixtures}: In polyphonic music, multiple notes are excited simultaneously. The observed acoustic signal is a linear mixture of vibrational modes produced by multiple strings, resulting in composite, intertwined spectro-temporal profiles.
    \item \textbf{Harmonic Interference}: Musical harmony relies heavily on consonant musical intervals (e.g., octaves, perfect fifths, perfect fourths), whose fundamental frequencies adhere to small integer ratios. As a consequence, the overtone series of simultaneously sounding notes heavily overlap in the frequency domain, obscuring individual fundamental frequencies and complicating source separation \citep{emiya2010multipitch, vincent2010adaptive}.
    \item \textbf{Temporal Dynamics and Acoustic Reverberation}: Real-world piano performance exhibits complex temporal non-stationarity, including percussive hammer strikes, sustain pedal resonances, string coupling, and room reverberation. The decay phase of preceding notes lingers across subsequent frames, frequently violating assumptions of temporal independence \citep{benetos2013multiple, sigtia2016end}.
    \item \textbf{Acute Class Imbalance}: A modern standard acoustic piano spans 88 keys. Even in virtuosic performances, human pianists rarely strike more than 8 to 10 keys simultaneously. Consequently, in any given analysis frame, only a very small fraction of the 88 pitches are active, resulting in an extreme sparsity ratio of active to inactive notes (empirically measured at approximately 1:41 in benchmark training corpora). Standard classification objectives heavily bias predictors toward inactive predictions unless explicitly countered.
\end{enumerate}

\begin{figure}[t]
    \centering
    \includegraphics[width=0.7\textwidth]{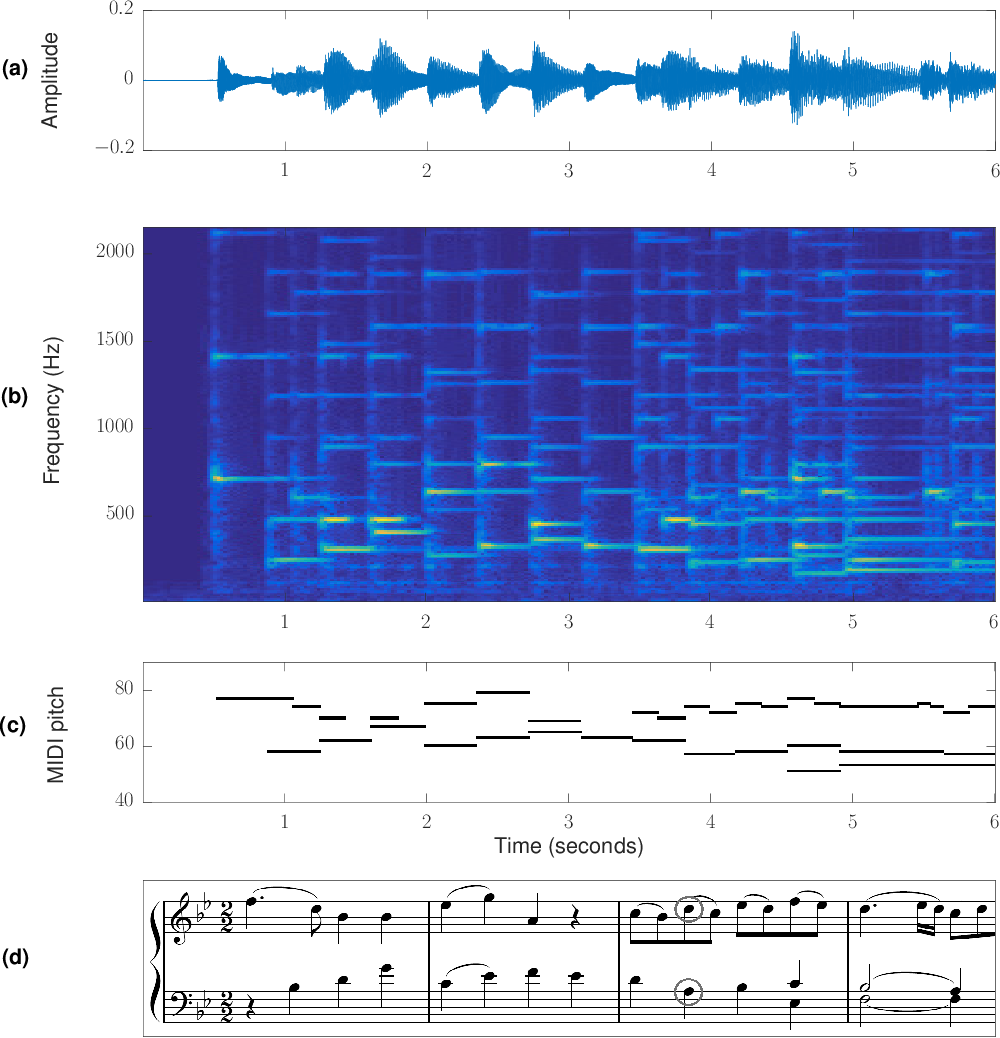}
    \caption{\textbf{Data representation pipeline in an Automatic Music Transcription (AMT) system.} The input acoustic waveform is transformed into a time-frequency representation, parameterized into frame-level feature vectors, processed by a sequence-modeling acoustic and temporal network, and decoded into active pitch activations (MIDI/piano roll representation) \citep{benetos2018automatic}.}
    \label{fig:amt_pipeline}
\end{figure}

To address these hurdles, this paper presents a learning-based framework for polyphonic piano transcription. As illustrated in Figure~\ref{fig:amt_pipeline}, we formulate the problem as a frame-level multi-label classification task over the 88 canonical piano pitches (MIDI notes 21 through 108). Recognizing the logarithmic frequency organization of musical pitch perception, we employ the Constant-Q Transform (CQT) \citep{brown1991calculation} over 7 octaves with 36 bins per octave, yielding a 252-dimensional spectral feature vector per time frame.

To model the temporal dynamics of music, we explore recurrent neural network architectures, contrasting unidirectional Long Short-Term Memory (LSTM) \citep{hochreiter1997long} networks against Bidirectional LSTMs (BiLSTM). Furthermore, we propose two novel attention extensions designed to operate harmoniously with recurrent temporal representations: a \textit{Modified Attention Mechanism} that incorporates explicit bias terms and hyperbolic tangent activation, and a \textit{Modified Multihead Attention} mechanism utilizing parallel attention projections combined with element-wise summation and non-linear compression. To address the 1:41 class imbalance, we formulate a Weighted Binary Cross-Entropy (WBCE) loss that penalizes false negatives on active notes. 

Through extensive experiments on the MAPS \citep{emiya2010maps} and MAESTRO \citep{hawthorne2018maestro} benchmark datasets, we systematically evaluate 7 distinct model configurations. Our best proposed model---a Multihead Attention--Bidirectional LSTM network---achieves an F1-score of 88.20\% and an overall accuracy of 78.90\% with only 4.43M trainable parameters, establishing an effective, lightweight solution for polyphonic music transcription.

\section{Related Work}
\label{sec:related_work}

\subsection{Classical Signal Processing and Matrix Factorization}
Research in automated music analysis traces back to the pioneering works of Moorer \citep{moorer1975, moorer1977}, who explored computer-based sound segmentation and transcription of continuous musical signals. Goto and Muraoka \citep{goto1994} subsequently pioneered real-time beat tracking and rhythm parsing from acoustic mixtures. Over the subsequent decades, multi-pitch estimation (MPE) focused on hand-crafted spectral representations and unsupervised decomposition. Prominent paradigms included non-negative matrix factorization (NMF) and its extensions \citep{smaragdis2003nonnegative, vincent2010adaptive, benetos2013multiple}, probabilistic spectral smoothness models \citep{emiya2010multipitch}, multiple fundamental frequency estimation via spectral peak/non-peak classification \citep{duan2010multiple}, and Bayesian generative spectrogram factorizations \citep{peeling2010generative, fuentes2013harmonic}. While these methods offered sound mathematical interpretability, they struggled to generalize to expressive dynamics and dense polyphony without rigid acoustic priors.

\subsection{Supervised Machine Learning and Deep Neural Networks}
The advent of supervised discriminative learning marked a paradigm shift in AMT. Poliner and Ellis \citep{poliner2007discriminative} deployed an ensemble of 87 binary Support Vector Machines (SVMs) followed by a Hidden Markov Model (HMM) for temporal smoothing, demonstrating that learning pitch-specific decision boundaries from data outperformed hand-crafted acoustic heuristics. 

With the deep learning resurgence, neural architectures rapidly advanced AMT performance. Sigtia et al. \citep{sigtia2016end, sigtia2015arxiv} introduced an end-to-end neural framework combining a convolutional neural network (CNN) acoustic front-end with an RNN language model to capture note transition semantics. Kelz et al. \citep{kelz2016potential} provided an influential study on simple framewise CNN approaches, demonstrating that high-capacity convolutional models could achieve competitive framewise transcription without explicit language modeling. Ullrich and van der Wel \citep{ullrich2017music} advanced sequence-to-sequence convolutional networks for polyphonic score estimation. 

Subsequent studies by Mor{\'a}n \citep{moran2017deep} and Li et al. \citep{li2017music} examined feed-forward Deep Neural Networks (DNN) and LSTM networks for framewise piano transcription, identifying the challenge of training stability and noting that simple recurrent networks often suffered from under-fitting or slow convergence when trained with mean squared error (MSE) loss and sub-optimal activation functions. Our work directly builds upon and resolves these limitations by introducing specialized attention modules, multi-label probabilistic loss functions, and bidirectional processing.

\subsection{Attention Mechanisms in Sequence Modeling}
Attention mechanisms, first introduced in machine translation by Bahdanau et al. \citep{bahdanau2015neural}, resolved the information bottleneck of fixed-size recurrent context vectors by enabling decoders to dynamically attend to relevant encoder hidden states across arbitrary temporal distances. Vaswani et al. \citep{vaswani2017attention} generalized this concept into the Multi-Head Attention and Transformer architectures, proving that multi-subspace projections allow models to jointly attend to information from different representation subspaces. In audio and music processing, attention mechanisms have become indispensable for capturing long-range harmonic and temporal correlations across disparate musical measures. In this paper, we adapt these mechanisms into modified recurrent attention blocks optimized for frame-level spectral inputs.

\section{Dataset and Preprocessing}
\label{sec:dataset}

\subsection{Dataset Description}
To evaluate our transcription models on polyphonic piano compositions across diverse acoustic conditions, we utilize two publicly accessible benchmark datasets:
\begin{itemize}
    \item \textbf{MAPS Database (MIDI Aligned Piano Sounds)} \citep{emiya2010maps}: MAPS contains high-fidelity audio recordings of classical piano pieces, chords, and isolated notes produced by upright and grand pianos (both physical instruments and high-grade software synthesizers). Each audio file is paired with ground truth text annotations specifying the exact onset time, offset time, and MIDI pitch for every note.
    \item \textbf{MAESTRO Dataset (MIDI and Audio Edited for Synchronous Tracks and Organization)} \citep{hawthorne2018maestro}: MAESTRO encompasses over 200 hours of virtuosic piano performances collected from ten years of the International Piano-e-Competition. The audio is captured from Yamaha Disklavier concert grands with sub-3\,ms alignment between MIDI events and acoustic waveforms.
\end{itemize}

The datasets are partitioned into training, validation, and test splits following standard evaluation protocols. As detailed in Table~\ref{tab:dataset_splits}, the MAPS dataset comprises 23,928 training files, 2,991 validation files, and 2,991 test files. The MAESTRO dataset provides 962 training compositions and 174 validation pieces.

\begin{table}[t]
  \caption{Data split distribution for the MAPS and MAESTRO benchmark datasets.}
  \label{tab:dataset_splits}
  \centering
  \begin{tabular}{lccc}
    \toprule
    \textbf{Dataset} & \textbf{Train} & \textbf{Validation} & \textbf{Test} \\
    \midrule
    MAESTRO \citep{hawthorne2018maestro} & 962 & 174 & -- \\
    MAPS \citep{emiya2010maps} & 23,928 & 2,991 & 2,991 \\
    \bottomrule
  \end{tabular}
\end{table}

\begin{table}[t]
  \caption{Sample extract of ground truth note event annotations pairing onset time, offset time, and MIDI pitch.}
  \label{tab:label_format}
  \centering
  \begin{tabular}{ccc}
    \toprule
    \textbf{Onset Time (s)} & \textbf{Offset Time (s)} & \textbf{MIDI Pitch} \\
    \midrule
    0.988281 & 1.111979 & 34 (A\#1) \\
    1.109375 & 1.135417 & 53 (F3) \\
    1.109375 & 1.161458 & 62 (D4) \\
    1.153646 & 1.196615 & 58 (A\#3) \\
    1.518229 & 1.549479 & 74 (D5) \\
    1.527344 & 1.563802 & 62 (D4) \\
    1.533854 & 1.566406 & 58 (A\#3) \\
    1.539062 & 1.566406 & 82 (A\#5) \\
    1.518229 & 1.567708 & 86 (D6) \\
    1.540365 & 1.578125 & 53 (F3) \\
    \bottomrule
  \end{tabular}
\end{table}

\begin{figure}[t]
    \centering
    \includegraphics[width=0.85\textwidth]{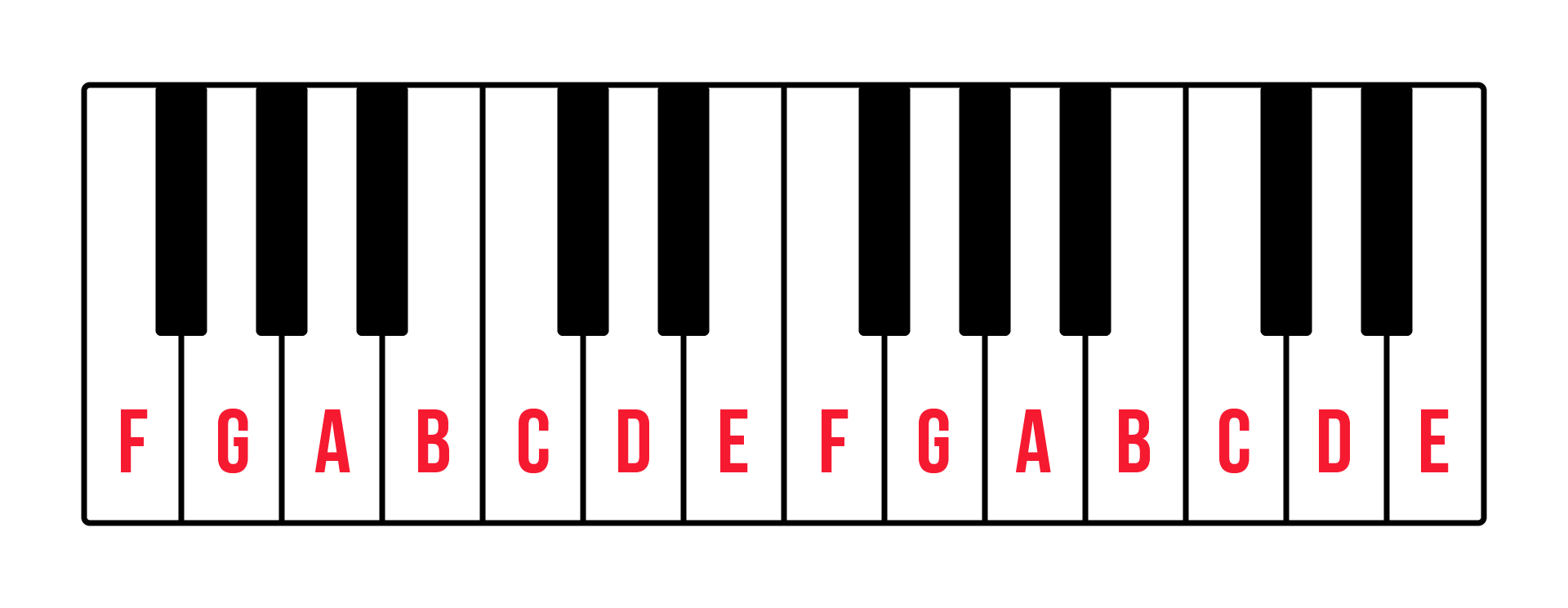}
    \caption{\textbf{Piano keyboard layout spanning 88 keys and 7 complete octaves.} Fundamental frequencies range from A0 (27.5\,Hz) to C8 (4186.01\,Hz), geometrically spaced according to the equal-tempered scale.}
    \label{fig:piano_notes}
\end{figure}

\subsection{Audio Preprocessing and Constant-Q Transform}
Raw audio waveforms stored in uncompressed \texttt{.wav} format are standardized to a sampling rate of $f_s = 44.1$\,kHz. Multichannel signals are down-mixed to mono by computing the channel-wise arithmetic mean:
\begin{equation}
    x[n] = \frac{1}{C}\sum_{c=1}^C x_c[n].
\end{equation}

While the Short-Time Fourier Transform (STFT) and Mel-Frequency Cepstral Coefficients (MFCC) are popular in general audio processing, they possess notable shortcomings for music transcription. The STFT enforces a uniform frequency resolution $\Delta f$ across the entire spectrum, which produces insufficient frequency resolution at low frequencies (where adjacent piano semitones differ by only a few Hertz) and redundant resolution at high frequencies. Conversely, MFCCs discard fine harmonic partial structures vital for multi-pitch disambiguation.

\begin{figure}[t]
    \centering
    \includegraphics[width=0.9\textwidth]{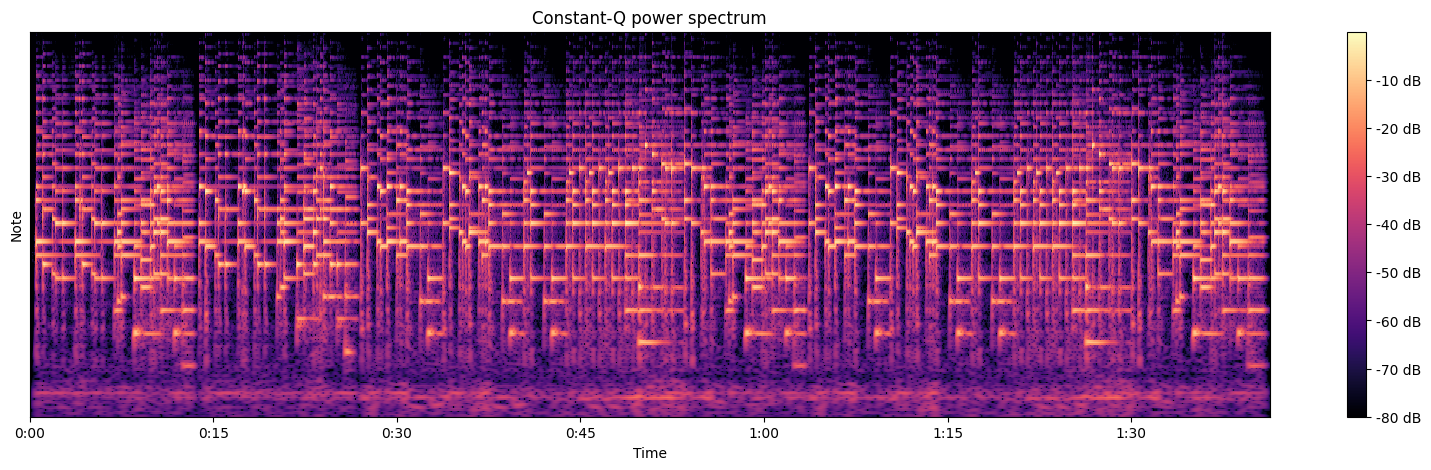}
    \caption{\textbf{Constant-Q Transform (CQT) spectrogram.} The log-frequency axis maintains constant musical interval spacing across octaves, providing sharp resolution for fundamental frequencies in the lower register.}
    \label{fig:cqt_spectrum}
\end{figure}

To align the spectral analysis with musical acoustics, we compute the Constant-Q Transform (CQT) \citep{brown1991calculation}. The CQT employs geometrically spaced frequency bins where the ratio of center frequency $f_k$ to bandwidth $\Delta f_k$ remains constant:
\begin{equation}
    Q = \frac{f_k}{\Delta f_k} = \frac{1}{2^{1/b} - 1},
\end{equation}
where $b$ denotes the number of bins per octave. In our implementation, the CQT is evaluated across $7$ octaves with $b = 36$ bins per octave (yielding 3 bins per semitone for pitch nuance) and a hop size of $H = 512$ samples (corresponding to a temporal resolution of $\Delta t \approx 11.6$\,ms per frame). The resulting spectral feature vector at each time frame $t$ comprises:
\begin{equation}
    K = 7 \times 36 = 252 \text{ frequency bins}.
\end{equation}
The continuous spectrogram is depicted in Figure~\ref{fig:cqt_spectrum}. Audio frames are grouped into continuous chunks of $40,000$ frames to facilitate uniform batch tensor operations during training.

\begin{figure}[t]
    \centering
    \begin{subfigure}[b]{0.32\textwidth}
        \centering
        \includegraphics[width=\textwidth]{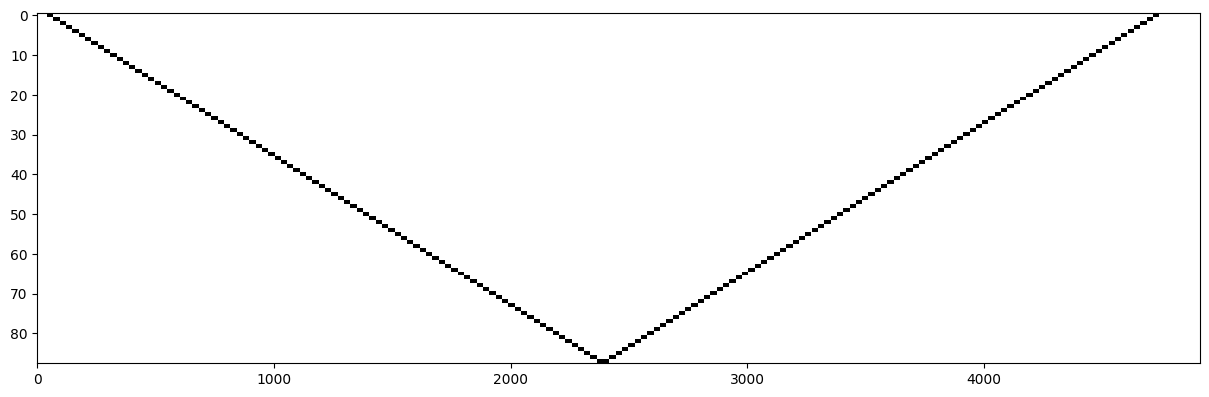}
        \caption{Sparse pitch activations}
    \end{subfigure}
    \hfill
    \begin{subfigure}[b]{0.32\textwidth}
        \centering
        \includegraphics[width=\textwidth]{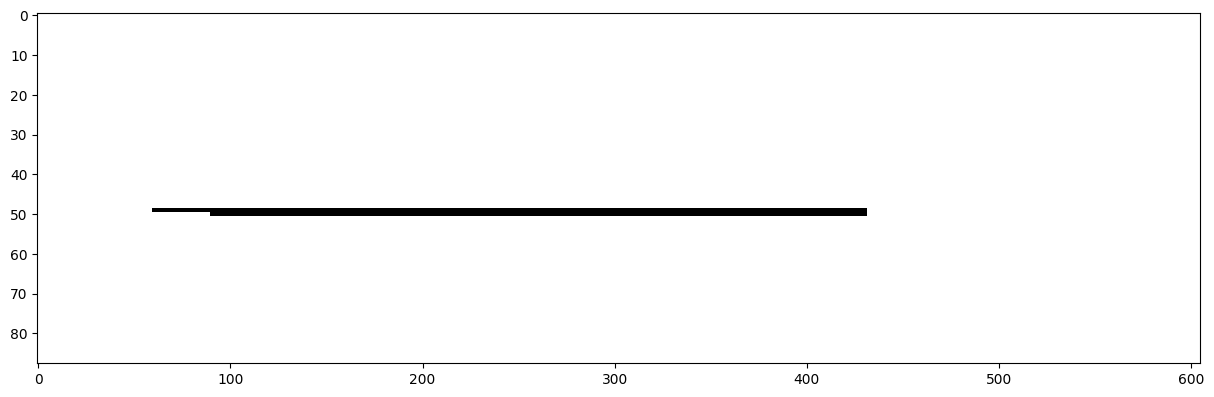}
        \caption{Polyphonic chord events}
    \end{subfigure}
    \hfill
    \begin{subfigure}[b]{0.32\textwidth}
        \centering
        \includegraphics[width=\textwidth]{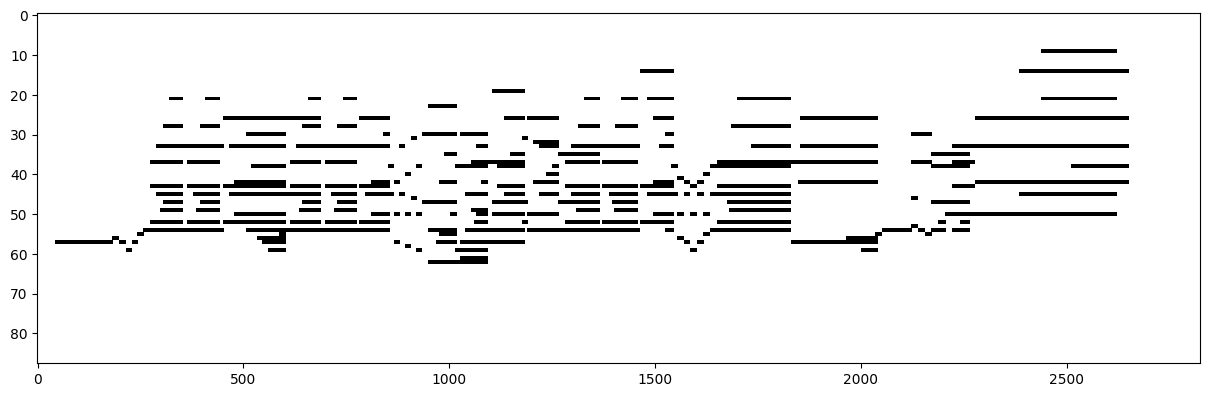}
        \caption{Sustained note transitions}
    \end{subfigure}
    \caption{\textbf{Visual illustration of frame-level multi-label ground truth matrices $\mathbf{y}_t \in \{0, 1\}^{88}$.} Rows correspond to the 88 piano pitches and columns represent consecutive time frames.}
    \label{fig:labels_example}
\end{figure}

\begin{figure}[t]
    \centering
    \includegraphics[width=0.45\textwidth]{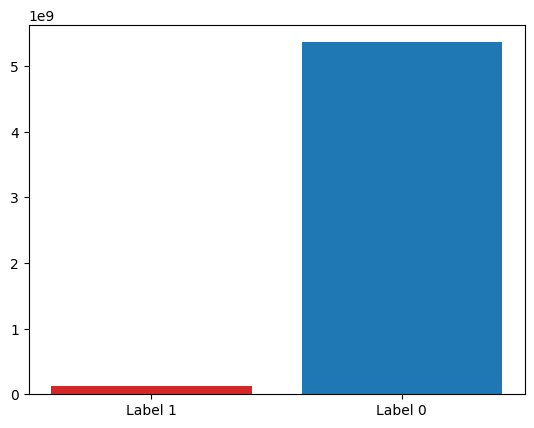}
    \caption{\textbf{Class distribution imbalance in the training corpus.} Inactive pitch labels (0) total 5.36 billion instances, whereas active pitch labels (1) total 0.129 billion instances, establishing a severe ratio of approximately 41:1.}
    \label{fig:class_imbalance}
\end{figure}

\subsection{Ground Truth Label Formulation and Data Imbalance}
The ground truth text files report musical events characterized by triplets: $\langle \text{onset}, \text{offset}, \text{pitch} \rangle$, as exemplified in Table~\ref{tab:label_format}. Because a standard piano comprises 88 keys spanning MIDI pitch numbers $21$ (A0, 27.5\,Hz) to $108$ (C8, 4186.01\,Hz) (Figure~\ref{fig:piano_notes}), we formulate the transcription target as a frame-level binary vector:
\begin{equation}
    \mathbf{y}_t = [y_{t, 1}, y_{t, 2}, \dots, y_{t, 88}]^T \in \{0, 1\}^{88},
\end{equation}
where $y_{t, p} = 1$ indicates that pitch $p$ is sounding at time frame $t$, and $y_{t, p} = 0$ otherwise. Stacking these vectors across time produces the binary piano-roll representation shown in Figure~\ref{fig:labels_example}.

\noindent \textbf{Data Imbalance Insight}: Statistical auditing of the compiled training corpus reveals an acute class imbalance. As shown in Figure~\ref{fig:class_imbalance}, the aggregate count of inactive labels ($y_{t,p} = 0$) reaches $5.36 \times 10^9$, whereas active notes ($y_{t,p} = 1$) account for only $0.129 \times 10^9$ entries. The resulting ratio of negative-to-positive instances is approximately:
\begin{equation}
    \text{Ratio} \approx \frac{5.36 \times 10^9}{0.129 \times 10^9} \approx 41.5 : 1.
\end{equation}
If left unaddressed, standard neural models trained with unweighted cross-entropy converge toward a trivial local optimum that predicts zeros across all pitches, yielding superficially high unweighted accuracy but near-zero recall. This empirical insight directly motivates our loss formulation in Section~\ref{sec:method}.

\section{Proposed Methodology}
\label{sec:method}

\subsection{Mathematical Formulation of the Constant-Q Transform}
The discrete Constant-Q Transform $X[k]$ of a discrete-time signal $x[n]$ is formulated as:
\begin{equation}
    X[k] = \frac{1}{N_k} \sum_{n=0}^{N_k-1} W[k, n]\, x[n]\, \exp\left(-j 2\pi Q \frac{n}{N_k}\right),
\end{equation}
where $k \in \{0, 1, \dots, K-1\}$ denotes the frequency bin index, $W[k, n]$ is a window function (e.g., Hamming window) of length $N_k$, and the window length $N_k$ varies inversely with the bin center frequency $f_k$:
\begin{equation}
    N_k = Q \frac{f_s}{f_k}.
\end{equation}
By scaling $N_k$ inversely with frequency, lower octaves utilize longer temporal windows to achieve fine spectral separation, while higher octaves utilize shorter windows to capture rapid transient percussive attacks.

\subsection{Recurrent Base Networks}

\subsubsection{Long Short-Term Memory (LSTM)}
Standard Recurrent Neural Networks (RNNs) maintain a hidden state vector $\mathbf{h}_t = \tanh(\mathbf{W}_h \mathbf{h}_{t-1} + \mathbf{W}_x \mathbf{x}_t + \mathbf{b})$. However, when backpropagated over extended sequences, gradient magnitudes vanish or explode exponentially \citep{hochreiter1997long}, preventing vanilla RNNs from learning long-term musical dependencies such as cadential closures or sustained notes.

\begin{figure}[t]
    \centering
    \includegraphics[width=0.7\textwidth]{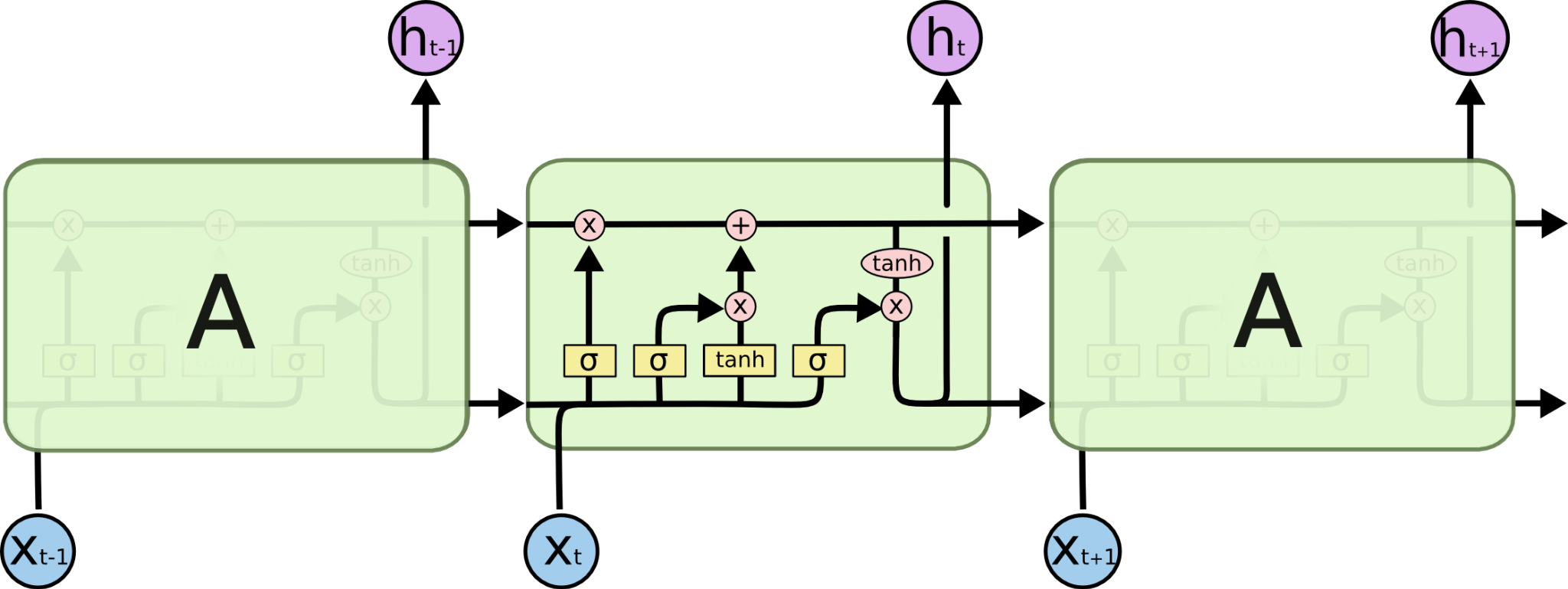}
    \caption{\textbf{Internal architecture of the Long Short-Term Memory (LSTM) cell.} The cell state $\mathbf{c}_t$ serves as a constant-error conveyor belt regulated by the forget gate $\mathbf{f}_t$, input gate $\mathbf{i}_t$, and output gate $\mathbf{o}_t$ \citep{hochreiter1997long}.}
    \label{fig:lstm_cell}
\end{figure}

The LSTM architecture overcomes this by introducing an explicit cell state vector $\mathbf{c}_t$ regulated by three multiplicative gating mechanisms (Figure~\ref{fig:lstm_cell}). At time step $t$, given input feature vector $\mathbf{x}_t \in \mathbb{R}^{252}$ and previous hidden state $\mathbf{h}_{t-1} \in \mathbb{R}^{d}$, the forward computations are:
\begin{align}
    \mathbf{f}_t &= \sigma(\mathbf{W}_f [\mathbf{h}_{t-1}, \mathbf{x}_t] + \mathbf{b}_f), \label{eq:forget_gate} \\
    \mathbf{i}_t &= \sigma(\mathbf{W}_i [\mathbf{h}_{t-1}, \mathbf{x}_t] + \mathbf{b}_i), \label{eq:input_gate} \\
    \tilde{\mathbf{c}}_t &= \tanh(\mathbf{W}_c [\mathbf{h}_{t-1}, \mathbf{x}_t] + \mathbf{b}_c), \label{eq:candidate_state} \\
    \mathbf{c}_t &= \mathbf{f}_t \odot \mathbf{c}_{t-1} + \mathbf{i}_t \odot \tilde{\mathbf{c}}_t, \label{eq:cell_update} \\
    \mathbf{o}_t &= \sigma(\mathbf{W}_o [\mathbf{h}_{t-1}, \mathbf{x}_t] + \mathbf{b}_o), \label{eq:output_gate} \\
    \mathbf{h}_t &= \mathbf{o}_t \odot \tanh(\mathbf{c}_t), \label{eq:hidden_update}
\end{align}
where $\sigma(z) = (1 + e^{-z})^{-1}$ denotes the logistic sigmoid activation, $\odot$ represents the Hadamard (element-wise) product, and $[\mathbf{h}_{t-1}, \mathbf{x}_t]$ denotes tensor concatenation.

\subsubsection{Bidirectional LSTM (BiLSTM)}
\begin{figure}[t]
    \centering
    \includegraphics[width=0.6\textwidth]{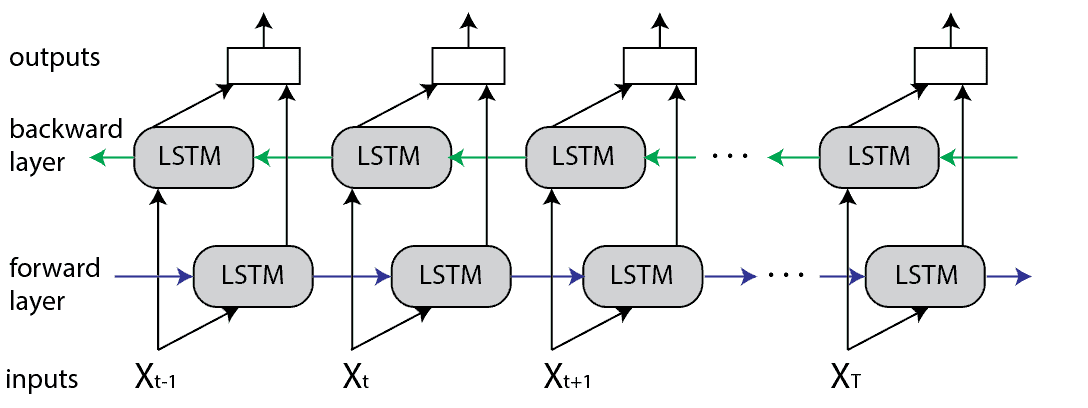}
    \caption{\textbf{Bidirectional Long Short-Term Memory architecture.} The input sequence is processed concurrently in both chronological (forward) and reverse (backward) temporal directions.}
    \label{fig:bilstm_arch}
\end{figure}

Unidirectional LSTMs process musical signals strictly chronologically ($t = 1 \to T$). In musical compositions, however, the harmonic function of a note is frequently clarified by its subsequent resolution (e.g., a dominant seventh chord resolving to a tonic chord). Bidirectional LSTMs (BiLSTM) process the sequence in both forward and backward directions using two independent hidden layers (Figure~\ref{fig:bilstm_arch}):
\begin{align}
    \overrightarrow{\mathbf{h}}_t &= \text{LSTM}_{\text{fwd}}(\mathbf{x}_t, \overrightarrow{\mathbf{h}}_{t-1}), \\
    \overleftarrow{\mathbf{h}}_t &= \text{LSTM}_{\text{bwd}}(\mathbf{x}_t, \overleftarrow{\mathbf{h}}_{t+1}).
\end{align}
The combined representation at frame $t$ is formed by concatenation:
\begin{equation}
    \mathbf{h}_t = [\overrightarrow{\mathbf{h}}_t, \overleftarrow{\mathbf{h}}_t] \in \mathbb{R}^{2d}.
\end{equation}
The aggregated probability vector $\mathbf{p}_t$ combines the forward and backward predictions:
\begin{equation}
    \mathbf{p}_t = \mathbf{p}_t^f + \mathbf{p}_t^b.
\end{equation}

\subsection{Attention Mechanisms for Music Transcription}

\subsubsection{Baseline Additive Attention (Bahdanau)}
The classical additive attention mechanism \citep{bahdanau2015neural} computes alignment scores between encoder hidden states $\mathbf{h}_i$ and the decoder query state $\mathbf{s}_{t-1}$:
\begin{equation}
    e_{t, i} = a(\mathbf{s}_{t-1}, \mathbf{h}_i) = \mathbf{v}_a^T \tanh(\mathbf{W}_a \mathbf{s}_{t-1} + \mathbf{U}_a \mathbf{h}_i),
\end{equation}
where $\mathbf{v}_a^T, \mathbf{W}_a$, and $\mathbf{U}_a$ are learnable projection matrices. Normalized attention weights $\alpha_{t, i}$ are obtained via the softmax operator:
\begin{equation}
    \alpha_{t, i} = \frac{\exp(e_{t, i})}{\sum_{j=1}^T \exp(e_{t, j})},
\end{equation}
and the context vector $\mathbf{c}_t$ is computed as the weighted sum:
\begin{equation}
    \mathbf{c}_t = \sum_{i=1}^T \alpha_{t, i} \mathbf{h}_i.
\end{equation}

\subsubsection{Proposed Modified Attention Mechanism}
Standard additive attention was originally formulated for sequence-to-sequence text generation with variable input/output lengths. In frame-level AMT, input spectral frames and output prediction vectors are temporally synchronized. To tailor attention to this dense setting, we introduce a bias parameter $\mathbf{b}_i$ and wrap the full alignment function within a bounded hyperbolic tangent operator:
\begin{equation}
    e_{t, i} = \tanh(a(\mathbf{s}_{t-1}, \mathbf{h}_i) + \mathbf{b}_i).
    \label{eq:modified_attention_score}
\end{equation}
The bounded range $(-1, 1)$ prevents extreme logit divergences in frame-level spectral attention. Attention weights are computed via softmax:
\begin{equation}
    \boldsymbol{\alpha}_t = \text{Softmax}(\mathbf{e}_t),
\end{equation}
and the resulting context vector is obtained by modulating the hidden state:
\begin{equation}
    \mathbf{c}_i = \alpha_{t, i}\, \mathbf{h}_i.
\end{equation}

\begin{figure}[t]
    \centering
    \includegraphics[width=0.45\textwidth]{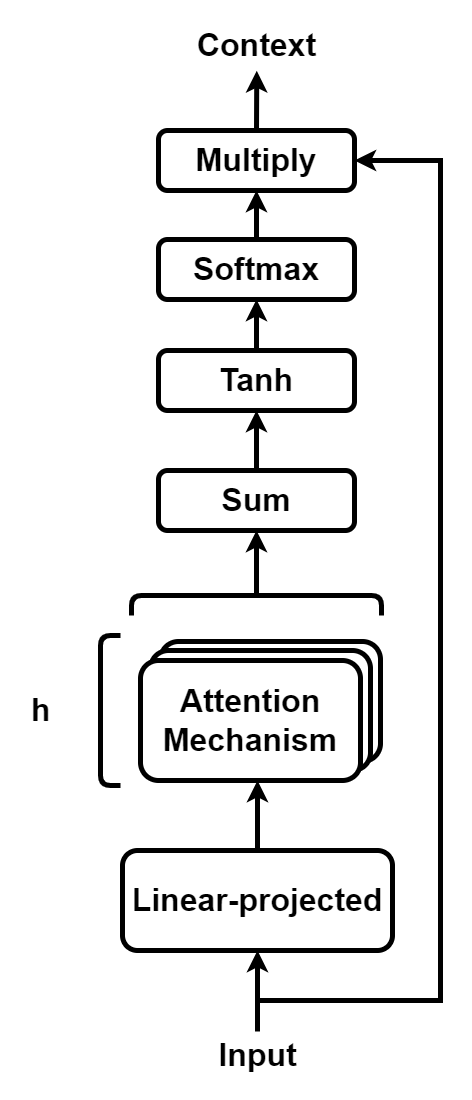}
    \caption{\textbf{Proposed Modified Multihead Attention module.} Parallel attention heads compute subspace alignment scores, which are summed, projected via $\mathbf{W}^O$ with bias $\mathbf{b}^O$, and activated by hyperbolic tangent before softmax normalization.}
    \label{fig:modified_multihead}
\end{figure}

\subsubsection{Proposed Modified Multihead Attention Mechanism}
In Transformer architectures \citep{vaswani2017attention}, standard Multi-Head Attention computes scaled dot-product attention over $K$ parallel projections:
\begin{equation}
    \text{Attention}(\mathbf{Q}, \mathbf{K}, \mathbf{V}) = \text{Softmax}\left(\frac{\mathbf{Q} \mathbf{K}^T}{\sqrt{d_k}}\right) \mathbf{V},
\end{equation}
followed by concatenation:
\begin{equation}
    \text{MultiHead}(\mathbf{Q}, \mathbf{K}, \mathbf{V}) = \text{Concat}(\text{head}_1, \dots, \text{head}_K) \mathbf{W}^O.
\end{equation}

To integrate multihead capacity into recurrent latent representations without excessive parameter inflation, we propose the \textit{Modified Multihead Attention} module illustrated in Figure~\ref{fig:modified_multihead}. Instead of scaled dot-product operations, each head computes alignment using our modified attention formulation (Equation~\ref{eq:modified_attention_score}). Rather than concatenating heads, the head outputs are aggregated via summation, transformed by a projection matrix $\mathbf{W}^O$ and bias $\mathbf{b}^O$, and passed through a hyperbolic tangent activation:
\begin{equation}
    \text{MultiHead}(e_{t, i}) = \tanh\left( \left(\sum_{k=1}^K \text{head}_k\right) \mathbf{W}^O + \mathbf{b}^O \right).
    \label{eq:mod_multihead_eq}
\end{equation}
The final attention weights and context vectors are derived as:
\begin{align}
    \boldsymbol{\alpha}_t &= \text{Softmax}(\text{MultiHead}(e_{t, i})), \\
    \mathbf{c}_i &= \alpha_{t, i}\, \mathbf{h}_i.
\end{align}
This structure enables the network to attend to complementary spectral sub-bands and temporal dependencies across multiple projection subspaces simultaneously.

\subsection{Activation and Loss Functions}

\subsubsection{Activation Functions}
For hidden layers and attention projections, the hyperbolic tangent activation function is utilized:
\begin{equation}
    \tanh(z) = \frac{e^z - e^{-z}}{e^z + e^{-z}} = \frac{1 - e^{-2z}}{1 + e^{-2z}},
\end{equation}
mapping activations to $(-1, 1)$. For the final transcription output layer, each of the 88 pitch nodes employs the independent sigmoid activation function:
\begin{equation}
    \hat{y}_{t, p} = \sigma(z_{t, p}) = \frac{1}{1 + e^{-z_{t, p}}}, \quad p \in \{1, 2, \dots, 88\},
\end{equation}
where $\hat{y}_{t, p} \in (0, 1)$ denotes the predicted probability that pitch $p$ is sounding at time frame $t$. Unlike the softmax function (which imposes a categorical probability distribution summing to 1 and assumes monophony), independent sigmoid activations naturally accommodate arbitrary polyphonic note combinations.

\subsubsection{Binary Cross-Entropy (BCE)}
Standard Binary Cross-Entropy treats all pitch predictions uniformly across frames:
\begin{equation}
    \mathcal{L}_{\text{BCE}} = -\frac{1}{N} \sum_{i=1}^N \Big[ y_i \log(\hat{y}_i) + (1 - y_i) \log(1 - \hat{y}_i) \Big].
\end{equation}

\subsubsection{Weighted Binary Cross-Entropy (WBCE)}
Given the severe 41:1 class imbalance documented in Section~\ref{sec:dataset}, standard BCE penalizes false positives and false negatives symmetrically, causing models to favor all-zero outputs. To counteract this, we employ Weighted Binary Cross-Entropy (WBCE):
\begin{equation}
    \mathcal{L}_{\text{WBCE}} = -\frac{1}{N} \sum_{i=1}^N \Big[ w_1\, y_i \log(\hat{y}_i) + w_0\, (1 - y_i) \log(1 - \hat{y}_i) \Big],
    \label{eq:wbce}
\end{equation}
where $w_1$ and $w_0$ represent positive (active note) and negative (inactive note) weighting parameters. In our benchmark experiments, we configure $w_1 = 0.65$ and $w_0 = 0.35$, placing higher penalty on missed notes to drive aggressive recall recovery.

\subsection{Post-Processing and Evaluation Metrics}
\label{sec:metrics}

\subsubsection{Post-Processing}
Raw model probabilities $\hat{y}_{t, p}$ are binarized using a threshold $\tau = 0.5$. To refine the frame-level output into clean symbolic notes, we apply a lightweight morphological post-processing filter:
\begin{enumerate}
    \item \textbf{Minimum Duration Thresholding}: Active note segments shorter than a minimum threshold (e.g., $< 3$ frames, $\approx 35$\,ms) are eliminated as spurious acoustic noise.
    \item \textbf{Gap Filling}: Gaps of 1 to 2 frames between consecutive activations of the same pitch are bridged to eliminate false note-off events caused by temporary harmonic dips.
\end{enumerate}

\subsubsection{Evaluation Metrics}
Model transcription accuracy is quantified using frame-level Precision, Recall, Accuracy, and F1-score:
\begin{align}
    \text{Precision} &= \frac{\text{TP}}{\text{TP} + \text{FP}}, \\
    \text{Recall} &= \frac{\text{TP}}{\text{TP} + \text{FN}}, \\
    \text{Accuracy} &= \frac{\text{TP}}{\text{TP} + \text{FP} + \text{FN}}, \\
    \text{F1-score} &= \frac{2 \times \text{Precision} \times \text{Recall}}{\text{Precision} + \text{Recall}},
\end{align}
where $\text{TP}, \text{FP}$, and $\text{FN}$ denote the aggregate counts of True Positives, False Positives, and False Negatives, respectively, computed across all 88 pitches over the test corpus.

\section{Experimental Configurations and Architectures}
\label{sec:experiments}

\begin{figure}[t]
    \centering
    \begin{subfigure}[b]{0.48\textwidth}
        \centering
        \includegraphics[height=5.2cm]{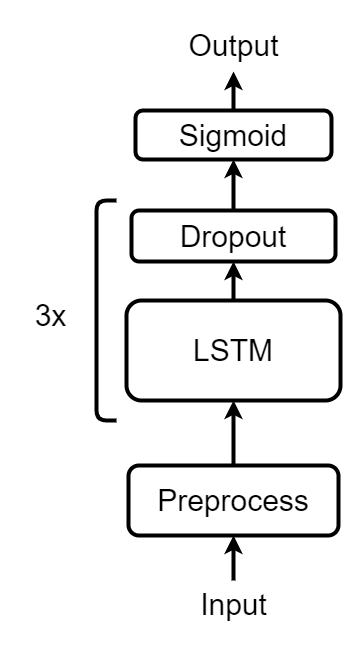}
        \caption{Baseline LSTM \citep{moran2017deep}}
    \end{subfigure}
    \hfill
    \begin{subfigure}[b]{0.48\textwidth}
        \centering
        \includegraphics[height=5.2cm]{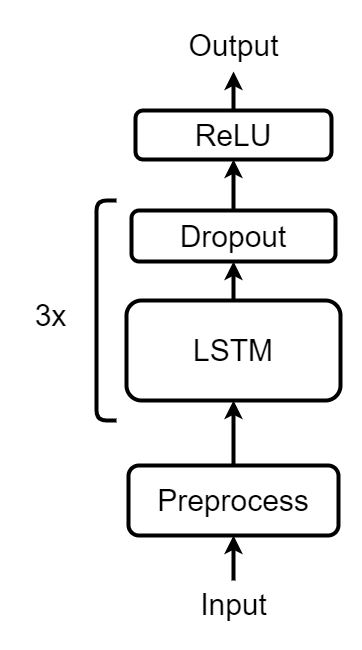}
        \caption{Modified Base LSTM (Ours)}
    \end{subfigure}
    \caption{\textbf{Architectural transition from baseline to modified base model.} Replacing MSE loss with WBCE and ReLU with Sigmoid eliminates saturation and enables effective multi-label learning.}
    \label{fig:base_models}
\end{figure}

To systematically dissect the contribution of each architectural innovation, we implement and benchmark seven sequential model configurations under identical training protocols:

\subsection{Model 1: Base Unidirectional LSTM}
Following the baseline setup of Mor{\'a}n \citep{moran2017deep}, the architecture consists of 3 stacked unidirectional LSTM layers with 256 hidden units per layer, each followed by a Dropout layer with rate $p = 0.20$. In contrast to \citep{moran2017deep} (which used MSE loss and ReLU outputs), our modified baseline adopts Sigmoid output activations and WBCE loss (Figure~\ref{fig:base_models}). Total parameters: 1.59M.

\subsection{Model 2: Bidirectional LSTM (BiLSTM)}
The 3 unidirectional LSTM layers are replaced with 3 Bidirectional LSTM layers (Figure~\ref{fig:model_architectures_a}). Each BiLSTM layer contains 256 forward units and 256 backward units (512 total hidden units per layer), followed by Dropout ($p = 0.20$). Total parameters: 4.24M.

\subsection{Model 3: LSTM--Attention}
Following the architecture of Bahdanau et al. \citep{bahdanau2015neural}, an attention layer is placed between two recurrent stages: an initial LSTM layer (256 units), our Modified Attention module, and two subsequent LSTM layers (256 units each) with Dropout ($p = 0.20$) (Figure~\ref{fig:model_architectures_b}). Total parameters: 1.66M.

\subsection{Model 4: Bidirectional LSTM--Attention}
\begin{figure}[p]
    \centering
    \begin{subfigure}[b]{0.31\textwidth}
        \centering
        \includegraphics[height=7.2cm]{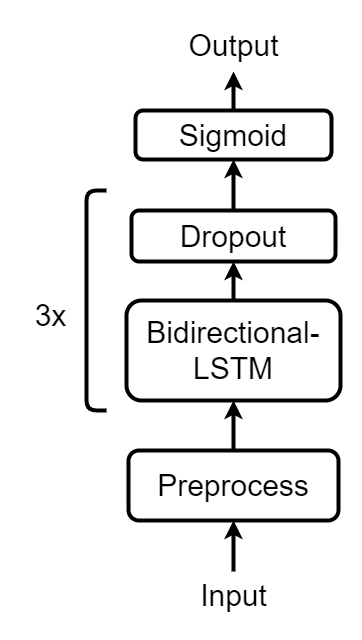}
        \caption{Bidirectional LSTM}
        \label{fig:model_architectures_a}
    \end{subfigure}
    \hfill
    \begin{subfigure}[b]{0.31\textwidth}
        \centering
        \includegraphics[height=7.2cm]{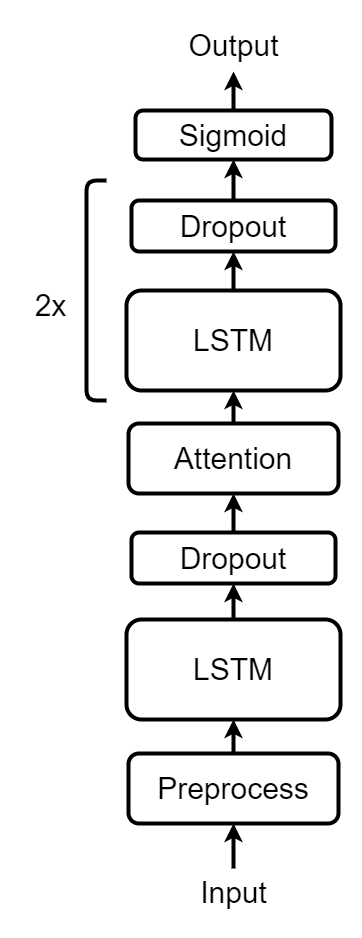}
        \caption{LSTM--Attention}
        \label{fig:model_architectures_b}
    \end{subfigure}
    \hfill
    \begin{subfigure}[b]{0.31\textwidth}
        \centering
        \includegraphics[height=7.2cm]{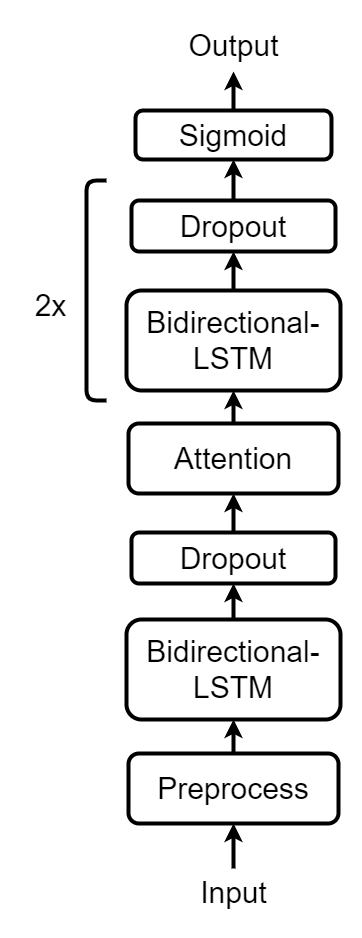}
        \caption{BiLSTM--Attention}
        \label{fig:model_architectures_c}
    \end{subfigure}
    
    \vspace{0.4cm}
    
    \begin{subfigure}[b]{0.31\textwidth}
        \centering
        \includegraphics[height=7.2cm]{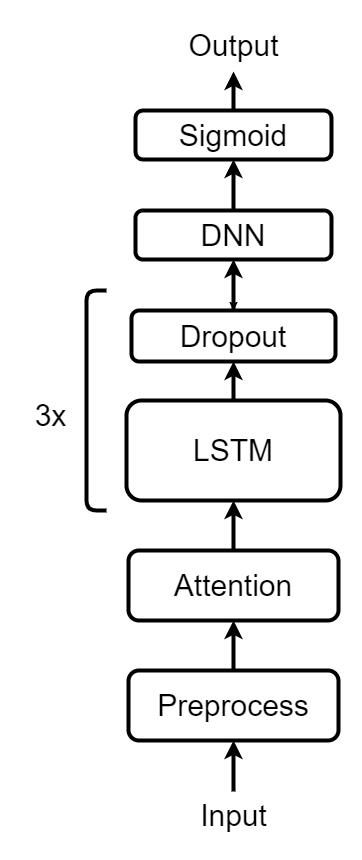}
        \caption{Attention--LSTM}
        \label{fig:model_architectures_d}
    \end{subfigure}
    \hfill
    \begin{subfigure}[b]{0.31\textwidth}
        \centering
        \includegraphics[height=7.2cm]{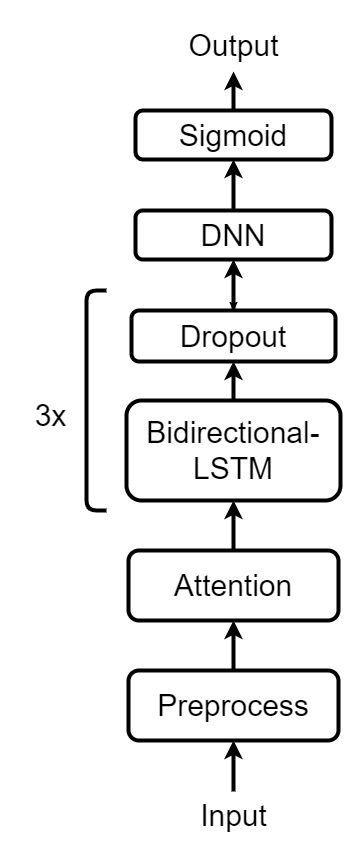}
        \caption{Attention--BiLSTM}
        \label{fig:model_architectures_e}
    \end{subfigure}
    \hfill
    \begin{subfigure}[b]{0.35\textwidth}
        \centering
        \includegraphics[height=7.2cm]{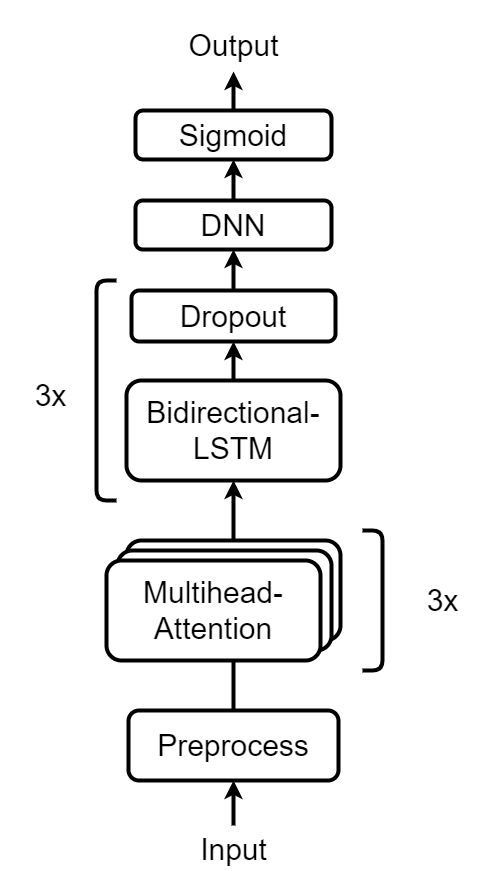}
        \caption{Multihead Attention--BiLSTM}
        \label{fig:model_architectures_f}
    \end{subfigure}
    \caption{\textbf{Systematic architectural configurations investigated in this study.} (a) Bidirectional LSTM; (b) Interleaved LSTM--Attention; (c) Interleaved BiLSTM--Attention; (d) Input Attention--LSTM; (e) Input Attention--BiLSTM; (f) Proposed Multihead Attention--BiLSTM architecture.}
    \label{fig:all_architectures}
\end{figure}

Extending Model 3, all LSTM layers are converted to Bidirectional LSTM layers (512 units per layer) flanking the Modified Attention module. Dropout is set to $p = 0.20$ (Figure~\ref{fig:model_architectures_c}). Total parameters: 4.35M.

\subsection{Model 5: Attention--LSTM}
Inspired by the Transformer paradigm where attention operates at the front of processing, we position the Modified Attention module at the input stage directly over the CQT spectral representations. The attended representations are subsequently processed by 3 stacked LSTM layers (256 units each), followed by a dense linear projection layer prior to the Sigmoid classifier (Figure~\ref{fig:model_architectures_d}). Total parameters: 1.66M.

\subsection{Model 6: Attention--Bidirectional LSTM}
Retaining the front-end Attention placement from Model 5, the recurrent backbone is upgraded to 3 BiLSTM layers (512 units each) with Dropout ($p = 0.20$) and linear projection (Figure~\ref{fig:model_architectures_e}). Total parameters: 4.35M.

\subsection{Model 7: Multihead Attention--Bidirectional LSTM (Proposed Best Model)}
Replacing the single-head attention module of Model 6, we deploy our Modified Multihead Attention module (Equation~\ref{eq:mod_multihead_eq}) configured with $K = 3$ parallel attention heads. The multi-head context representations feed into 3 BiLSTM layers (512 units each), a dense linear layer, and the 88-dimensional Sigmoid output classifier (Figure~\ref{fig:model_architectures_f}). Total parameters: 4.43M.

\subsection{Training Protocol and Optimization}
All models are implemented in PyTorch and optimized using the Adam optimizer with an initial learning rate of $\eta = 10^{-3}$ and default momentum parameters ($\beta_1 = 0.9, \beta_2 = 0.999$). Training is monitored via validation loss with early stopping configured at a patience of 10 epochs to prevent over-fitting.

\section{Results and Comparative Analysis}
\label{sec:results}

\subsection{Quantitative Performance Comparison}
Table~\ref{tab:main_results} reports the quantitative transcription performance across all 7 evaluated models, along with ablation variants examining loss functions and training data diversity.

\begin{table}[t]
  \caption{\textbf{Comprehensive benchmark evaluation of transcription models.} Precision, Recall, Accuracy (\%), and F1-score (\%) on the polyphonic piano test dataset.}
  \label{tab:main_results}
  \centering
  \small
  \begin{tabular}{lccccc}
    \toprule
    \textbf{Model Architecture} & \textbf{Parameters} & \textbf{Precision} & \textbf{Recall} & \textbf{Accuracy (\%)} & \textbf{F1-score (\%)} \\
    \midrule
    LSTM (Base Model) \citep{moran2017deep} & 1.59M & 0.76 & 0.66 & 55.20 & 71.20 \\
    Bidirectional LSTM & 4.24M & 0.77 & 0.74 & 61.20 & 75.98 \\
    LSTM--Attention & 1.66M & 0.74 & 0.75 & 59.80 & 74.80 \\
    Bidirectional LSTM--Attention & 4.35M & 0.80 & 0.83 & 69.50 & 82.00 \\
    Attention--LSTM & 1.66M & 0.80 & 0.84 & 69.80 & 82.20 \\
    Attention--Bidirectional LSTM & 4.35M & 0.84 & 0.88 & 75.50 & 86.10 \\
    Multihead--Bidirectional LSTM (WBCE) & 4.43M & 0.84 & \textbf{0.90} & 77.30 & 87.20 \\
    \midrule
    \textbf{Multihead--BiLSTM + BCE (Best Model)} & 4.43M & \textbf{0.89} & 0.86 & \textbf{78.90} & \textbf{88.20} \\
    Multihead--BiLSTM (Single Dataset: MAPS only) & 4.43M & 0.83 & 0.87 & 74.50 & 85.40 \\
    \bottomrule
  \end{tabular}
\end{table}

Several key findings emerge from the empirical results:
\begin{enumerate}
    \item \textbf{Impact of Bidirectional Context}: Transitioning from unidirectional LSTM to Bidirectional LSTM produces an immediate gain of $+4.78\%$ in F1-score (from $71.20\%$ to $75.98\%$) and $+6.00\%$ in accuracy. This confirms that backward temporal context is critical for resolving pitch ambiguities.
    \item \textbf{Attention Placement Matters}: Placing the attention module at the input front-end (Attention--LSTM, F1: $82.20\%$) substantially outperforms placing it between recurrent layers (LSTM--Attention, F1: $74.80\%$), yielding a $+7.40\%$ margin. Input-level attention enables the model to selectively filter relevant spectral bins prior to recurrent temporal integration.
    \item \textbf{Synergy of Multihead Attention and BiLSTM}: Combining front-end multihead attention with BiLSTM layers achieves an F1-score of $87.20\%$ under WBCE. The 3 attention heads independently attend to low-register fundamentals, mid-range chord harmonies, and upper partials.
    \item \textbf{Loss Function Tuning}: When using the Multihead--BiLSTM architecture, standard BCE achieves the highest overall accuracy ($78.90\%$) and F1-score ($88.20\%$), improving precision from $0.84$ to $0.89$. While WBCE maximizes recall ($0.90$) by heavily penalizing false negatives, it introduces a slight increase in false positive detections, whereas BCE provides a superior precision-recall balance once the model capacity is sufficiently high.
    \item \textbf{Dataset Diversity}: Restricting training exclusively to the MAPS dataset reduces F1-score to $85.40\%$ (a $2.80\%$ drop), demonstrating that co-training on MAPS and MAESTRO enhances generalizability across diverse piano timbres and recording environments.
\end{enumerate}

\begin{figure}[t]
    \centering
    \begin{subfigure}[b]{0.48\textwidth}
        \centering
        \includegraphics[width=\textwidth]{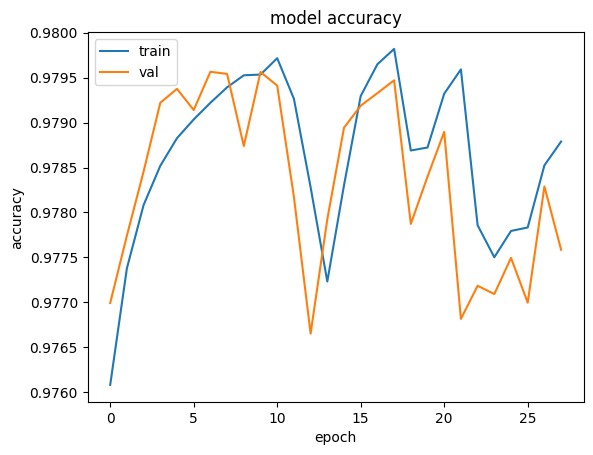}
        \caption{Base Model: Training Loss}
    \end{subfigure}
    \hfill
    \begin{subfigure}[b]{0.48\textwidth}
        \centering
        \includegraphics[width=\textwidth]{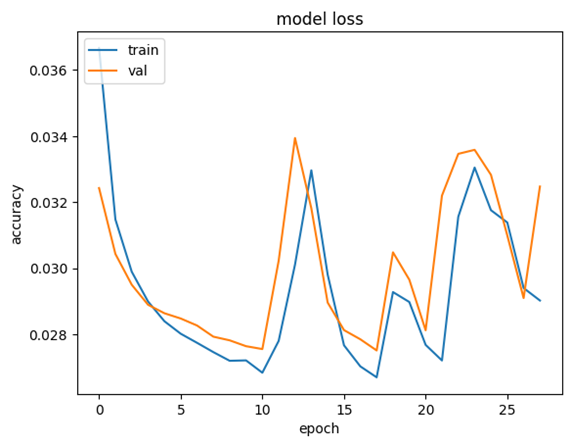}
        \caption{Base Model: Training Accuracy}
    \end{subfigure}
    
    \vspace{0.3cm}
    
    \begin{subfigure}[b]{0.48\textwidth}
        \centering
        \includegraphics[width=\textwidth]{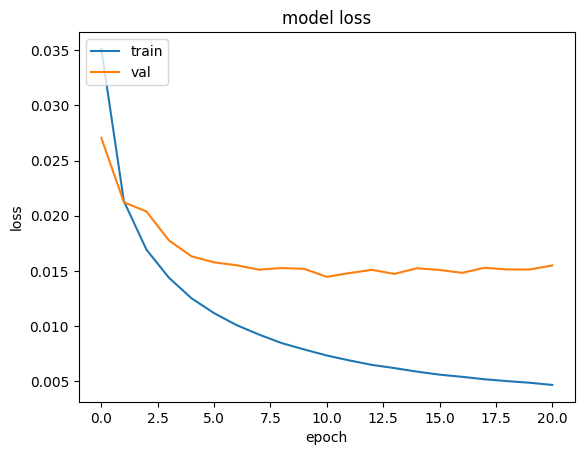}
        \caption{Best Model: Training Loss}
    \end{subfigure}
    \hfill
    \begin{subfigure}[b]{0.48\textwidth}
        \centering
        \includegraphics[width=\textwidth]{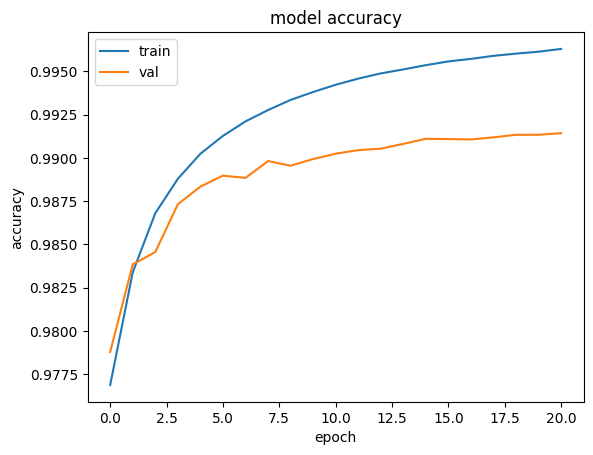}
        \caption{Best Model: Training Accuracy}
    \end{subfigure}
    \caption{\textbf{Training dynamics of Base Model (LSTM) versus Best Model (Multihead Attention--BiLSTM).} The proposed best model converges to a substantially lower training loss while maintaining rapid, steady accuracy improvements without over-fitting.}
    \label{fig:training_curves}
\end{figure}

\subsection{Training Dynamics}
Figure~\ref{fig:training_curves} illustrates the training loss and accuracy trajectories for the baseline LSTM versus the proposed Multihead Attention--BiLSTM. The baseline LSTM exhibits slower convergence and plateauing accuracy around $55\%$. In contrast, the Multihead Attention--BiLSTM demonstrates steep loss descent and steady accuracy escalation reaching nearly $80\%$, benefiting from stabilized gradient propagation via the tanh-bounded multihead attention module.

\begin{figure}[t]
    \centering
    \begin{subfigure}[b]{\textwidth}
        \centering
        \includegraphics[width=0.92\textwidth]{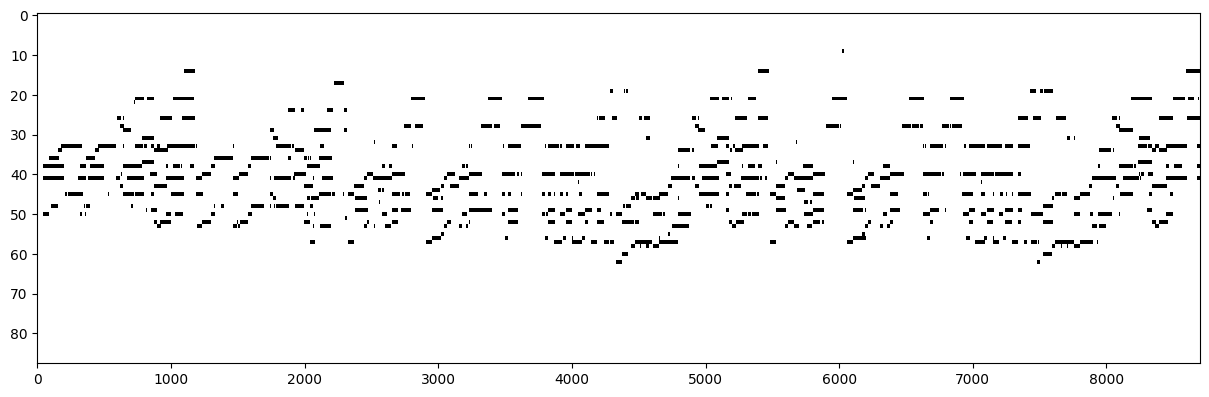}
        \caption{Predicted transcription: Base Model (LSTM)}
    \end{subfigure}
    
    \vspace{0.2cm}
    
    \begin{subfigure}[b]{\textwidth}
        \centering
        \includegraphics[width=0.92\textwidth]{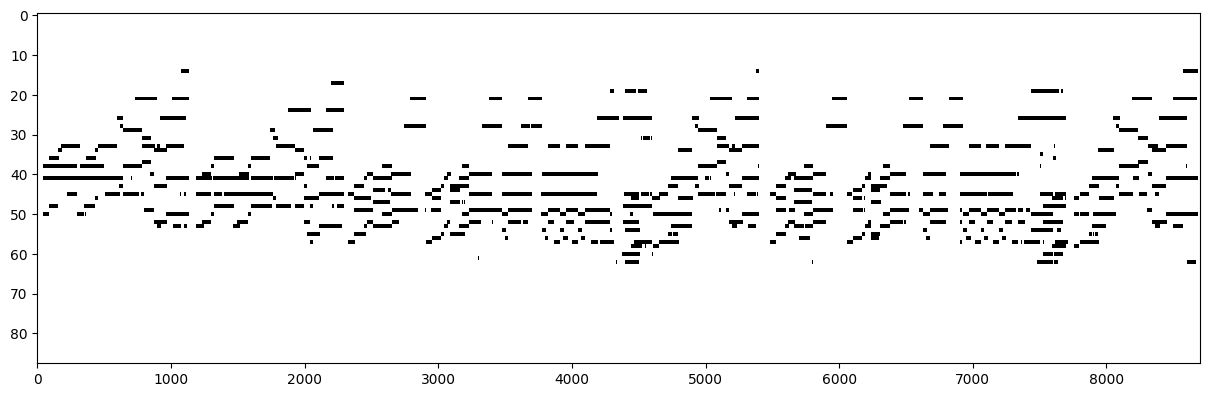}
        \caption{Predicted transcription: Best Model (Multihead Attention--BiLSTM)}
    \end{subfigure}
    
    \vspace{0.2cm}
    
    \begin{subfigure}[b]{\textwidth}
        \centering
        \includegraphics[width=0.92\textwidth]{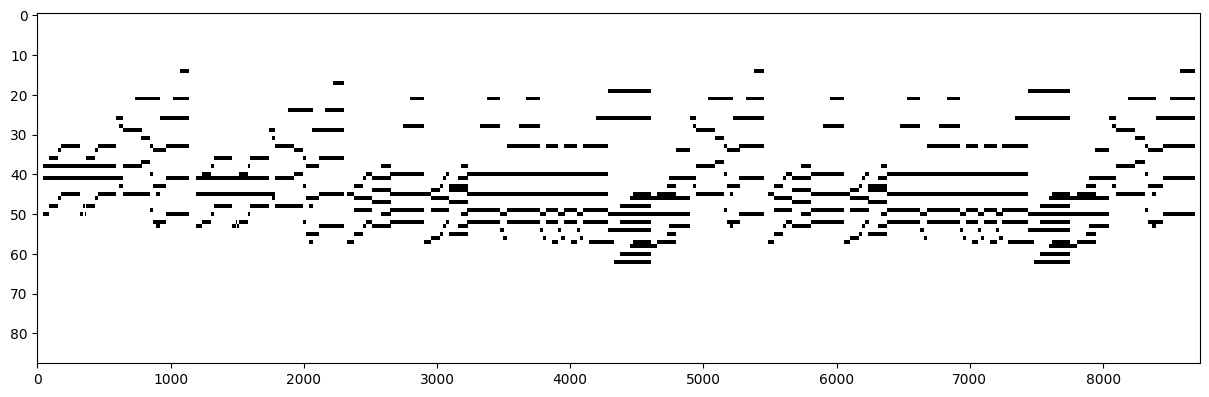}
        \caption{Ground Truth piano-roll annotation}
    \end{subfigure}
    \caption{\textbf{Visual comparison of transcribed piano rolls against Ground Truth.} The Base Model exhibits substantial temporal fragmentation and missed notes, whereas our Best Model closely reconstructs note onsets, sustained durations, and polyphonic chords.}
    \label{fig:prediction_comparison}
\end{figure}

\subsection{Qualitative Piano-Roll Visualizations}
Figure~\ref{fig:prediction_comparison} displays qualitative transcription predictions against the ground truth piano roll. The baseline LSTM (Figure~\ref{fig:prediction_comparison}a) suffers from substantial temporal fragmentation, spurious dropouts in sustained notes, and missed pitches in dense polyphonic intervals. In contrast, the predictions generated by our Multihead Attention--BiLSTM model (Figure~\ref{fig:prediction_comparison}b) closely match the ground truth (Figure~\ref{fig:prediction_comparison}c), preserving continuous pitch contours and precisely capturing polyphonic chord onsets.

\begin{figure}[t]
    \centering
    \begin{subfigure}[b]{\textwidth}
        \centering
        \includegraphics[width=0.92\textwidth]{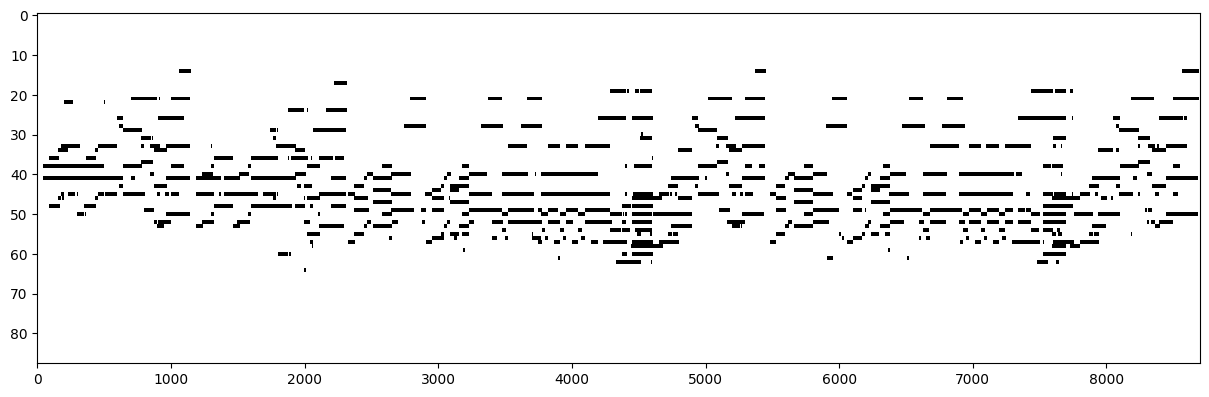}
        \caption{Prediction: Multihead--BiLSTM with Weighted Binary Cross-Entropy (WBCE)}
    \end{subfigure}
    
    \vspace{0.2cm}
    
    \begin{subfigure}[b]{\textwidth}
        \centering
        \includegraphics[width=0.92\textwidth]{image37.png}
        \caption{Prediction: Multihead--BiLSTM with standard Binary Cross-Entropy (BCE)}
    \end{subfigure}
    
    \vspace{0.2cm}
    
    \begin{subfigure}[b]{\textwidth}
        \centering
        \includegraphics[width=0.92\textwidth]{image38.png}
        \caption{Ground Truth annotation}
    \end{subfigure}
    \caption{\textbf{Qualitative ablation of loss formulation: WBCE versus standard BCE.} WBCE promotes higher sensitivity but introduces occasional background noise, whereas BCE achieves cleaner background suppression with high harmonic fidelity.}
    \label{fig:loss_comparison}
\end{figure}

\begin{figure}[t]
    \centering
    \begin{subfigure}[b]{\textwidth}
        \centering
        \includegraphics[width=0.92\textwidth]{image39.png}
        \caption{Prediction: Multihead--BiLSTM trained on Combined Datasets (MAPS + MAESTRO)}
    \end{subfigure}
    
    \vspace{0.2cm}
    
    \begin{subfigure}[b]{\textwidth}
        \centering
        \includegraphics[width=0.92\textwidth]{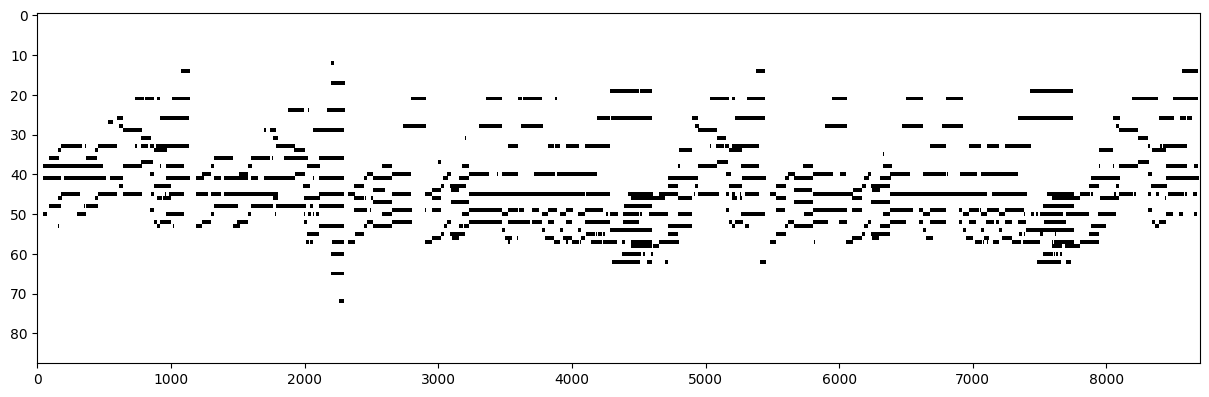}
        \caption{Prediction: Multihead--BiLSTM trained on Single Dataset (MAPS only)}
    \end{subfigure}
    
    \vspace{0.2cm}
    
    \begin{subfigure}[b]{\textwidth}
        \centering
        \includegraphics[width=0.92\textwidth]{image38.png}
        \caption{Ground Truth annotation}
    \end{subfigure}
    \caption{\textbf{Qualitative ablation of dataset diversity: Combined Datasets versus Single Dataset.} Training on both MAPS and MAESTRO provides broader acoustic exposure, reducing false insertions in higher octaves.}
    \label{fig:dataset_comparison}
\end{figure}

\subsection{Ablation Studies}

\subsubsection{Weighted BCE versus Standard BCE}
Figure~\ref{fig:loss_comparison} compares piano-roll outputs generated under WBCE versus standard BCE. While WBCE successfully prevents the network from collapsing into all-zero predictions during early training and maximizes recall ($0.90$), its higher positive weighting can induce faint ghost activations during soft reverberant decays. Standard BCE, when applied to our high-capacity attention architecture, effectively suppresses background noise while maintaining high precision ($0.89$), resulting in an optimal overall F1-score of $88.20\%$.

\subsubsection{Single Dataset versus Multi-Dataset Training}
Figure~\ref{fig:dataset_comparison} compares the model trained solely on MAPS against the model trained on both MAPS and MAESTRO. The single-dataset model exhibits sporadic false detections in the upper register, where overtone partials from lower octaves are mistakenly classified as independent notes. The multi-dataset model resolves these partial collisions more effectively, demonstrating the benefits of diverse training acoustics.

\section{Discussion and Limitations}
\label{sec:discussion}

While our proposed Multihead Attention--BiLSTM model achieves competitive transcription performance with a lightweight architecture ($4.43$M parameters), qualitative error analysis reveals several remaining challenges:
\begin{enumerate}
    \item \textbf{Dense Polyphony and Octave Collisions}: When multiple notes sharing exact harmonic intervals (e.g., octaves, fourths, fifths) sound concurrently, the fundamental frequency of the upper note coincides with a strong harmonic overtone of the lower note. In passages with high polyphonic density, the model occasionally misattributes lower partial energy to secondary notes or misses soft inner chord voices.
    \item \textbf{Tremolo and Rapid Articulations}: In tremolo passages where a performer rapidly repeats a single pitch, acoustic energy decays and builds within short intervals. Because CQT temporal integration and hop size introduce a frame resolution of $\approx 11.6$\,ms, rapid re-attacks can blur into a single continuous activation contour.
    \item \textbf{Up-Tempo Passages}: In virtuoso prestissimo compositions, note durations may span only 3 to 5 frames. Any minor misalignment in frame-level boundaries can exert a disproportionate impact on precision and recall metrics.
    \item \textbf{Heuristic Post-Processing}: Our current post-processing relies on fixed minimum-duration thresholds and small-gap filling. Incorporating a trainable language model or Hidden Markov Model (HMM) could provide context-aware temporal smoothing without manual threshold tuning.
\end{enumerate}

\section{Conclusion and Future Work}
\label{sec:conclusion}

In this paper, we presented a learning-based framework for polyphonic piano transcription that combines log-frequency Constant-Q Transform representations with Bidirectional LSTMs and specialized attention mechanisms. We designed a modified additive attention module and a modified multihead attention mechanism with hyperbolic tangent compression tailored to recurrent hidden states. Furthermore, we formulated a Weighted Binary Cross-Entropy loss to counteract the acute 41:1 negative-to-positive class imbalance. Systematic experiments across seven model variants on the MAPS and MAESTRO benchmarks proved that front-end attention combined with bidirectional temporal modeling achieves strong transcription accuracy: our Multihead Attention--BiLSTM model reached an F1-score of 88.20\% and an accuracy of 78.90\% with only 4.43M parameters.

Future work will explore hybrid convolutional-recurrent architectures (CRNNs) to learn adaptive low-level spectro-temporal filters directly from waveforms, multi-task learning incorporating joint onset, sustain, and velocity prediction heads, and transformer-based autoregressive music language models for score refinement.

\section*{Acknowledgments}
We express sincere gratitude to Mr.~Nguyen Quoc Trung and Mr.~Le Phu Nguyen for their constructive feedback and insightful technical reviews.

\section*{Author Contributions}
\textbf{Do Phu Anh Tai}: Conceptualization, lead research on recurrent deep learning architectures for AMT, and model development. \\
\textbf{Nguyen Huu Hoang}: Design and implementation of the modified attention and modified multihead attention mechanisms, and initial manuscript preparation. \\
\textbf{Nguyen Duc Hoang}: Development of the data preprocessing pipeline (CQT feature extraction, audio segmentation, multi-label annotation formatting), and manuscript review. \\
\textbf{Le Thanh Hai}: Research supervision, methodology refinement, project administration, and final manuscript editing. \\
All authors have reviewed and approved the final manuscript.

\section*{Conflicts of Interest}
The authors declare no conflicts of interest regarding the publication of this paper.

\bibliographystyle{unsrtnat}
\bibliography{references}

@phdthesis{moorer1975,
  title = {On the Segmentation and Analysis of Continuous Musical Sound by Digital Computer},
  author = {Moorer, James Anderson},
  school = {Stanford University},
  year = {1975}
}

@article{moorer1977,
  title = {On the transcription of musical sound by computer},
  author = {Moorer, James Anderson},
  journal = {Computer Music Journal},
  volume = {2},
  number = {1},
  pages = {7--11},
  year = {1977}
}

@inproceedings{goto1994,
  title = {A beat tracking system for acoustic signals of music},
  author = {Goto, Masataka and Muraoka, Yoichi},
  booktitle = {Proceedings of the Second ACM International Conference on Multimedia},
  pages = {365--372},
  year = {1994}
}

@techreport{klapuri2006,
  title = {Introduction to music transcription},
  author = {Klapuri, Anssi},
  institution = {Institute of Signal Processing, Tampere University of Technology},
  address = {Tampere, Finland},
  year = {2006}
}

@article{emiya2010multipitch,
  title = {Multipitch estimation of piano sounds using a new probabilistic spectral smoothness principle},
  author = {Emiya, Valentin and Badeau, Roland and David, Bertrand},
  journal = {IEEE Transactions on Audio, Speech, and Language Processing},
  volume = {18},
  number = {6},
  pages = {1643--1654},
  year = {2010}
}

@article{duan2010multiple,
  title = {Multiple fundamental frequency estimation by modeling spectral peaks and non-peak regions},
  author = {Duan, Zhiyao and Pardo, Bryan and Zhang, Changshui},
  journal = {IEEE Transactions on Audio, Speech, and Language Processing},
  volume = {18},
  number = {8},
  pages = {2121--2133},
  year = {2010}
}

@article{peeling2010generative,
  title = {Generative spectrogram factorization models for polyphonic piano transcription},
  author = {Peeling, Peter H. and Cemgil, Ali Taylan and Godsill, Simon J.},
  journal = {IEEE Transactions on Audio, Speech, and Language Processing},
  volume = {18},
  number = {3},
  pages = {519--527},
  year = {2010}
}

@inproceedings{smaragdis2003nonnegative,
  title = {Non-negative matrix factorization for polyphonic music transcription},
  author = {Smaragdis, Paris and Brown, Judith C.},
  booktitle = {IEEE Workshop on Applications of Signal Processing to Audio and Acoustics (WASPAA)},
  pages = {177--180},
  year = {2003}
}

@article{vincent2010adaptive,
  title = {Adaptive harmonic spectral decomposition for multiple pitch estimation},
  author = {Vincent, Emmanuel and Bertin, Nancy and Badeau, Roland},
  journal = {IEEE Transactions on Audio, Speech, and Language Processing},
  volume = {18},
  number = {3},
  pages = {528--537},
  year = {2010}
}

@article{benetos2013multiple,
  title = {Multiple-instrument polyphonic music transcription using a temporally-constrained shift-invariant model},
  author = {Benetos, Emmanouil and Dixon, Simon},
  journal = {The Journal of the Acoustical Society of America},
  volume = {133},
  number = {3},
  pages = {1727--1741},
  year = {2013}
}

@article{fuentes2013harmonic,
  title = {Harmonic adaptive latent component analysis of audio and application to music transcription},
  author = {Fuentes, Bernardo and Badeau, Roland and Richard, Ga{\"e}l},
  journal = {IEEE Transactions on Audio, Speech, and Language Processing},
  volume = {21},
  number = {9},
  pages = {1854--1866},
  year = {2013}
}

@article{sigtia2016end,
  title = {An end-to-end neural network for polyphonic piano music transcription},
  author = {Sigtia, Siddharth and Benetos, Emmanouil and Dixon, Simon},
  journal = {IEEE/ACM Transactions on Audio, Speech, and Language Processing},
  volume = {24},
  number = {5},
  pages = {927--939},
  year = {2016}
}

@inproceedings{kelz2016potential,
  title = {On the potential of simple framewise approaches to piano transcription},
  author = {Kelz, Rainer and Dorfer, Matthias and Korzeniowski, Filip and B{\"o}ck, Sebastian and Arzt, Andreas and Widmer, Gerhard},
  booktitle = {Proceedings of the 17th International Society for Music Information Retrieval Conference (ISMIR)},
  pages = {475--481},
  year = {2016}
}

@inproceedings{ullrich2017music,
  title = {Music transcription with convolutional sequence-to-sequence models},
  author = {Ullrich, Karen and van der Wel, Erik},
  booktitle = {Proceedings of the 18th International Society for Music Information Retrieval Conference (ISMIR)},
  pages = {464--471},
  year = {2017}
}

@article{poliner2007discriminative,
  title = {A discriminative model for polyphonic piano transcription},
  author = {Poliner, Graham E. and Ellis, Daniel P. W.},
  journal = {EURASIP Journal on Applied Signal Processing},
  volume = {2007},
  number = {1},
  pages = {154--154},
  year = {2007}
}

@misc{moran2017deep,
  title = {Deep neural networks for piano music transcription},
  author = {Gonz{\'a}lez Mor{\'a}n, Diego},
  howpublished = {\url{https://github.com/diegomorin8/Deep-Neural-Networks-for-Piano-Music-Transcription}},
  year = {2017}
}

@article{sigtia2015arxiv,
  title = {An end-to-end neural network for polyphonic piano music transcription},
  author = {Sigtia, Siddharth and Benetos, Emmanouil and Dixon, Simon},
  journal = {arXiv preprint arXiv:1508.01774},
  year = {2015}
}

@misc{li2017music,
  title = {Music Transcription Using Deep Learning},
  author = {Li, Luoqi and Ni, Isabella and Yang, Liang},
  howpublished = {Stanford University CS229 Final Project Report},
  year = {2017}
}

@inproceedings{bahdanau2015neural,
  title = {Neural machine translation by jointly learning to align and translate},
  author = {Bahdanau, Dzmitry and Cho, Kyunghyun and Bengio, Yoshua},
  booktitle = {International Conference on Learning Representations (ICLR)},
  year = {2015}
}

@article{brown1991calculation,
  title = {Calculation of a constant {Q} spectral transform},
  author = {Brown, Judith C.},
  journal = {The Journal of the Acoustical Society of America},
  volume = {89},
  number = {1},
  pages = {425--434},
  year = {1991}
}

@article{benetos2018automatic,
  title = {Automatic music transcription: An overview},
  author = {Benetos, Emmanouil and Dixon, Simon and Duan, Zhiyao and Ewert, Sebastian},
  journal = {IEEE Signal Processing Magazine},
  volume = {36},
  number = {1},
  pages = {20--30},
  year = {2018}
}

@inproceedings{vaswani2017attention,
  title = {Attention is all you need},
  author = {Vaswani, Ashish and Shazeer, Noam and Parmar, Niki and Uszkoreit, Jakob and Jones, Llion and Gomez, Aidan N. and Kaiser, {\L}ukasz and Polosukhin, Illia},
  booktitle = {Advances in Neural Information Processing Systems (NeurIPS)},
  volume = {30},
  pages = {5998--6008},
  year = {2017}
}

@misc{emiya2010maps,
  title = {{MAPS} Database: A piano database for multipitch estimation and automatic transcription of music},
  author = {Emiya, Valentin and Bertin, Nancy and David, Bertrand and Badeau, Roland},
  howpublished = {\url{https://adasp.telecom-paris.fr/resources/2010-07-08-maps-database/}},
  year = {2010}
}

@inproceedings{hawthorne2018maestro,
  title = {Enabling factorized piano music modeling and generation with the {MAESTRO} dataset},
  author = {Hawthorne, Curtis and Stasyuk, Andriy and Roberts, Adam and Simon, Ian and Huang, Cheng-Zhi Anna and Dieleman, Sander and Elsen, Erich and Engel, Jesse and Eck, Douglas},
  booktitle = {International Conference on Learning Representations (ICLR)},
  year = {2019}
}

@article{hochreiter1997long,
  title = {Long short-term memory},
  author = {Hochreiter, Sepp and Schmidhuber, J{\"u}rgen},
  journal = {Neural Computation},
  volume = {9},
  number = {8},
  pages = {1735--1780},
  year = {1997}
}

\end{document}